# Design Considerations of Biaxially Tensile-Strained Germanium-on-Silicon Lasers

Xiyue Li[a, b], Zhiqiang Li[c], Simon Li[c], Lukas Chrostowski[d], and Guangrui (Maggie) Xia[b]

[a]School of Electronic and Information Engineering, South China University of Technology, Guangzhou, Guangdong 510641, China

[b]Department of Materials Engineering, University of British Columbia, Vancouver, BC V6T 1Z4, Canada

[c]Crosslight Software Inc., Vancouver, BC V5M 2A4, Canada

[d]Department of Electrical and Computer Engineering, University of British Columbia, Vancouver, BC V6T 1Z4, Canada

Corresponding authors: Xiyue Li (email: lixiyue915@gmail.com), Guangrui(Maggie) Xia( email: gxia@mail.ubc.ca)

**Abstract**: Physical models of Ge energy band structure and material loss were implemented in LASTIP$^{TM}$, a 2D simulation tool for edge emitting laser diodes. This model is able to match experimental data available. Important design parameters of a Fabry-Perot Ge laser, such as the cavity length, thickness, width, polycrystalline Si cladding layer thickness were studied and optimized. The laser structure optimizations alone were shown to reduce $I_{th}$ by 22-fold and increase $\eta_d$ by 11 times. The simulations also showed that improving the defect limited carrier lifetime is critical for achieving an efficient and low-threshold Ge laser. With the optimized structure design (300 μm for the cavity length, 0.4 μm for the cavity width, 0.3 μm for the cavity thickness, and 0.6 μm for the polycrystalline Si cladding layer thickness) and a defect limited carrier lifetime of 100 ns, a wall-plug efficiency of 14.6% at 1mW output is predicted, where $J_{th}$ of 2.8 kA/cm$^2$, $I_{th}$ of 3.3 mA, $I_{1mW}$ of 9 mA, and $\eta_d$ of 23.6% can be achieved. These are tremendous improvements from the available experimental values at 280 kA/cm$^2$, 756 mA, 837 mA and 1.9%, respectively.

## 1. Background and Introduction

Germanium is an indirect bandgap semiconductor, which is inferior in light emitting applications compared to direct bandgap semiconductors, such as GaAs and InP. However, it is the most Si-compatible semiconductor and plays an important role in Si photonics, such as detectors [1] and modulators [2]. In the past few decades, researchers all over the world have made extensive efforts in finding solutions to a Si-compatible lasing material system [3]-[17].

Breakthroughs were made by a group of MIT researchers, who demonstrated that Ge can become a gain medium sufficient for laser applications by adding tensile strain [18] and heavy n-type doping [19]. In 2010, an optically pumped Ge-on-Si laser was demonstrated using 0.2% biaxial tensile strain [20]. It operated at room temperature with a gain of 50 cm$^{-1}$ at n-type doping of 1×10$^{19}$ cm$^{-3}$. The lasing was in a wavelength range of 1590 to 1610 nm. In 2012, an electrically pumped Ge-on-Si laser was demonstrated by the researchers from MIT and APIC Corporation, applying 4×10$^{19}$ cm$^{-3}$ n-type doping and 0.25% biaxial tensile strain [21]. The lasing wavelengths were between 1520 and 1700 nm with a variation consistent with different clamping condition. In 2015, lasing was observed from an electrically pumped 3×10$^{19}$ cm$^{-3}$ n-type doped Ge Fabry-Perot resonator on Si by R. Koerner et al. [22], confirming the principal validity of early work of Prof. Kimerling's group at MIT in Refs. [20, 21].

Introducing tensile strain to Ge is crucial to change Ge from an indirect bandgap material into a direct bandgap material [18, 21]. Both biaxial and uniaxial tensile strain can make this transition. Many efforts have been made to increase tensile strains in Ge. The standard one relied on the thermal expansion coefficient mismatch during the growth and cooling process of Ge on Si layer [18]. It can lead to approximately 0.25% biaxial tensile strain in Ge layers. G. Capellini *et al.* used silicon nitride layer to stress Ge up to about 0.9% biaxial tensile strain, and the fabrication process was COMS-compatible [23]. D. S. Sukhdeo *et al.* from Stanford used a stress concentration method in Ge-on-insulator (GOI) substrates, and obtained 5.7% uniaxial tensile stress in Ge bridges. Their work also showed that at 4.6% uniaxial tensile strain Ge changes to a direct bandgap material [24]. On the doping technologies, multi-layered delta-doped layers were used to create source phosphorous (P) concentration above $1\times 10^{20}$ cm$^{-3}$ and achieve active carrier concentration above $4\times 10^{19}$ cm$^{-3}$ [25, 26].

On the theoretical modeling side, the gain calculation model and a threshold current density model were described by Cai and Han et al. [27]. With the bandgap narrowing effect and the energy separation effect for heavily n-type doped Ge, good agreements with the experiment data can be obtained. W. W. Chow also described a gain theory for a bulk Ge active medium taking the many-body Coulomb effect into account [28]. B. Dutt et al. studied the Ge doping and strain impacts on slope efficiencies with some simplified assumptions without taking the cavity dimensions into account [29]. All the above studies are theoretical calculations without taking the laser dimensions or the light field into account.

As an early demonstration, the laser in Ref. [21] had an extremely high threshold current, which limited its operation efficiency. To shed some light on the laser design and performance improvements, two-dimensional (2D) device simulations are in great need. Little work is available on 2D laser device simulations, and key device performance parameters such as threshold current and wall-plug efficiencies have not been well-studied, which will be addressed in this work.

In this work, models of Ge energy band structure (with biaxial strain and doping) and material loss (including the bandgap narrowing effect and the energy separation effect) were implemented in LASTIP$^{TM}$, a 2D simulation tool for edge emitting laser diodes. With this capability, simulations of a Ge Fabry-Perot laser under biaxial tensile strain were made to investigate the important design parameters and to optimize the Ge laser performance.

## 2. Optical gain model and parameters used

The biaxial tensile strain impact on Ge energy bandgap structure has been well studied, and a biaxial tensile strain can be considered as a combination of a hydrostatic strain component plus a uniaxial strain component [30, 31]. The strain related Ge energy band models in Refs. [30, 31] was implemented in LASTIP$^{TM}$. Doping induced bandgap narrowing effect (BGN) is non-negligible [32]. The direct bandgap narrowing effect ($\Delta E_g^{\Gamma}$) model from Ref. [32] was implemented in LASTIP$^{TM}$, while the indirect bandgap narrowing effect $\Delta E_g^L$ was assumed the same as $\Delta E_g^{\Gamma}$ [27]. The effective mass values of electrons and holes in Ge we used are from Ref. [18].

Gain calculations were made for Ge with 0.25% biaxial tensile strain and $4\times 10^{19}$ cm$^{-3}$ n-type doping at different carrier injection levels, as shown in Fig. 1. The calculations are consistent with literature work in Ref. [27]. The kinks at about $\lambda = 1720$ nm show that the quasi-Fermi level $F_v$ has already entered the HH band inducing fast rise in gain coefficient due to significantly increased density of states.

In the low strain region ($\varepsilon_{xx} < 0.6\%$), the gain spectra maximum is associated with the direct band to band transition involving heavy hole band (HH band) that indeed hosts the majority of valence carrier [33], which means population inversion into HH band is required to overcome free carrier losses.

Direct transitions between $\Gamma_c$ and light hole (LH) states will occur for larger values of biaxial strain [29]. These favor that the Ge light emission under 0.25% tensile strain is operating in TE mode, which is relevant to the HH transition [33]. Additionally, for the Ge laser simulated in Ref. [21], the losses for TM modes are very large due to the metal contact, and only TE modes have low enough losses for light emission [26].

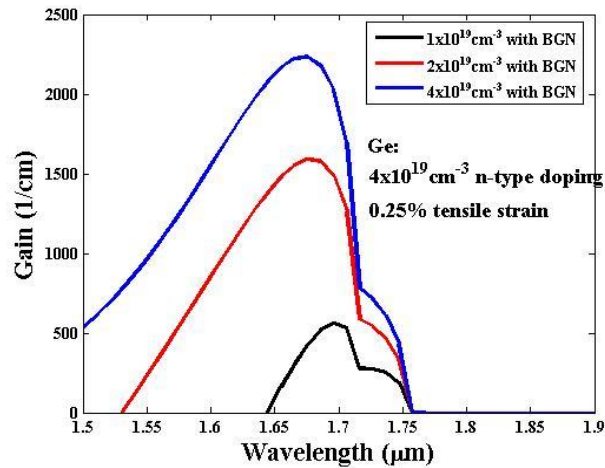

Fig.1. Gain spectra from direct transition in 0.25% tensile strained n$^+$ Ge with $4\times10^{19}$ cm$^{-3}$ doping at different carrier injection level.

## 3. Optimizations of an electrically pumped Fabry-Perot Germanium laser

The experimental work in Ref. [21] showed a lasing range from 1520 nm to 1700 nm, a threshold current density ($J_{th}$) of 280 kA/cm$^2$, and a differential quantum efficiency of 1.9%. The reason for this high $J_{th}$ is due to the non-optimized cavity dimensions, the fabrication imperfections including high series resistance from the top poly-Si contact, large free carrier absorption, metals contacts as well as the high diode leakage current [27]. Moreover, only 1 mW of output was observed despite the high current density of 310 kA/cm$^2$ was driven into the 270 μm long Ge cavity. The device cannot be practical without both drastic reduction in the threshold current and a vastly improved efficiency. Therefore, investigating methods to optimize and improve the Ge laser performance is the main target of this work. Although Ge lasers can also be used in long distance optical communications, the target application and thus the performance optimization of this work are for the on-chip optical interconnects. After the Ge cavity and cladding thickness optimization to reduce the threshold current ($I_{th}$), the effect of defect limited carrier time is also taken into account. No thermal effects are included in this work.

### 3.1 Laser structure, parameters to optimize and optimization criteria

In this work, we investigated the design optimization of an electrically pumped Ge-on-Si double heterojuction Fabry-Perot laser, and the cross section of this laser is shown in Fig. 2. The laser structure, the top metal contact, doping and strain are the same as the experiment reported in Ref. [21]. The bottom metal contact is not shown as it is sufficiently far away. Only 2 μm Si substrate is included in the simulations, which is set to be $5\times10^{19}$ cm$^{-3}$ n-type doped [26]. A virtual contact was defined underneath the bottom of the 2 μm Si substrate for biasing purpose and has no interactions with photons or the light intensity distribution.

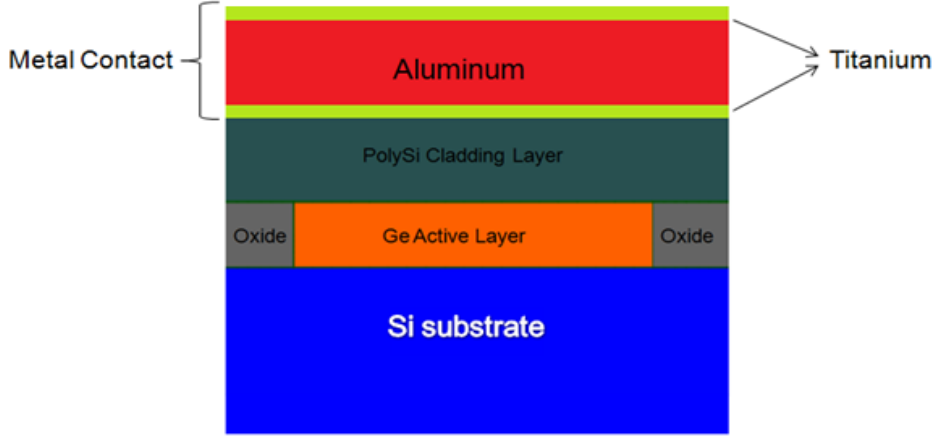

Fig. 2. Cross section of the Ge-on-Si heterojunction laser structure simulated.

The parameters to optimize are the Ge cavity length, thickness, width, and the poly-Si cladding thickness. In our optimization, five figures of merits are discussed. First, the threshold current density $J_{th}$ is used so that our simulation can be compared with available experimental data or simulations. For the Ge thickness and poly-Si thickness dependence study, as no cross-section area change is involved, $J_{th}$ and the threshold current $I_{th}$ share the same trend. Then, in the light power-current (L-I) plots, $I_{th}$ and the current at 1 mW optical output power ($I_{1mW}$) are quoted. The choice of 1 mW is just for the convenience of discussion. The differential quantum efficiency $\eta_d$ and the wall-plug efficiency $\eta_{wp}$ are calculated to benchmark the Ge laser efficiency in converting electrical power to light power. Among all the figures of merits, the most important optimization criterion is the threshold current $I_{th}$.

### 3.2 Parameters used and fitting the experimental data

For on-chip optical interconnect applications, $I_{th}$ is a very important figure of merit, and it is highly desirable to have $I_{th}$ as low as possible. The data available in experimental work Ref. [21] have a $J_{th}$ of 280 kA/cm$^2$, which corresponds to $I_{th}$ = 756 mA, and a differential quantum efficiency of 1.9%. We need to first calibrate our models with the experimental data. For that purpose, we set the structure, doping and stress parameters same as those in Ref. [21], which is under 0.25% biaxial tensile strain and $4\times10^{19}$ cm$^{-3}$ n-type doping, 1 μm Ge width, 270 μm length and 180 nm poly-Si cladding layer thickness. The Ge active layer thickness was set to be 200 nm, which is the average value of the 100~300 nm thickness in the experiment due to the non-uniform interface [21, 26]. The reflectively values of two facet are $R_1$ = 23% and $R_2$ = 38%, and the corresponding mirror loss $\alpha_m$ is 45 cm$^{-1}$ [27]. The index of refraction of Ge is set to be 4.2 [34]. Auger coefficients used were $C_{nnp}$ = 3.0× 10$^{-32}$ cm$^6$/s and $C_{ppn}$ = 7.0× 10$^{-32}$ cm$^6$/s [18]. The defect limited carrier lifetime in epitaxial grown Ge film is conservatively assumed to be 1 ns for this thickness based on measurements in recent Refs. [40], [41]. Those measurements were not performed on junction structures, where the field near the junctions could place the minority carriers away from the highly defected Ge/Si interfaces. Therefore, we consider the 1 ns value is a conservative estimation. In section 3.7, the carrier lifetime dependence is further discussed.

In this work, we assume that the internal loss and the mirror loss are the main sources of absorption, and the internal loss is dominated by the free carrier absorption [35]. In LASTIP$^{TM}$, for a narrow wavelength range, the free carrier absorption is described by

$$\alpha_i = AN + BP, \qquad (1)$$

where A and B are constants, N and P are electron and hole density in units of $cm^{-3}$.

The only fitting parameters we used to fit the experimental L-I characteristics were the parameters in the free carrier absorption of Ge in Equation (1). We used the first principle calculations of free carrier absorption results in n-type doped Ge [26, 36] and experimental measurements in p-type doped Ge [37] as a starting point and obtained the best fitting to the L-I characteristics with the following free carrier absorption:

$$\alpha_i = 5.0 \times 10^{-19}N + 1.03 \times 10^{-17}P \ . \tag{2}$$

The free carrier absorption relations for the doped Si substrate and poly-Si cladding layer are from well-known work in Refs. [38] and [39].

Using these parameters, our models produce a $J_{th}$ of 290 $kA/cm^2$ and a $I_{th}$ of 790 mA at 15℃ with the TE mode at λ = 1680 nm, which is relevant to the gain peak location in Fig. 1, very close to the experimental value of 280 $kA/cm^2$ [21]. As seen in Fig.3, the models can match the experimental L-I curve quite well. After the calibration of our models, we started optimizing the laser structure. The Ge length dependence was first studied.

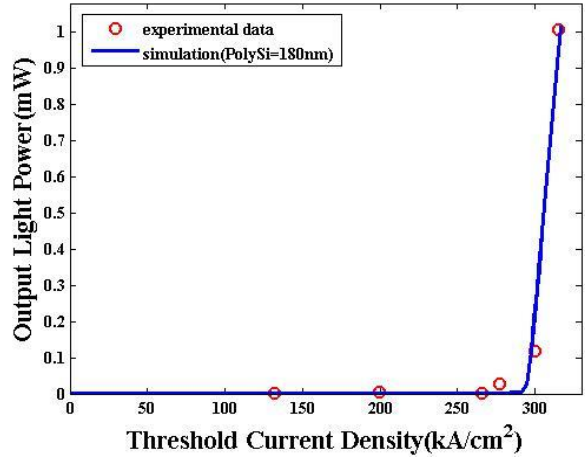

Fig.3. Comparisons between the results using our models and the experimental data from Ref. [21] at 15℃ (Ge length: 270μm, Ge thickness: 200nm, Ge width: 1μm, poly-Si thickness: 180 nm).

### 3.3 Ge cavity length dependence

To optimize the Ge cavity length, the L-I characteristics were simulated for a range of Ge cavity length from 270 μm to 1 mm with other parameters unchanged. Theoretically, the longer cavity can both generate more power and reduce the impact of the mirror loss. The cavity length of common compound semiconductor lasers can reach as long as 1 mm. However, larger threshold current is not desired for on-chip interconnects applications. Therefore, there still is an optimal cavity length for the lowest $I_{th}$. In our simulations, $I_{th}$ and $I_{1mW}$ can be obtained by the L-I plots. Just as it is desirable to have as low $I_{th}$ as possible, it is also desirable to have an efficient laser. From these two current values $I_{th}$ and $I_{1mW}$, we can deduct the differential quantum efficiency $\eta_d$ [42],

$$\eta_d = \frac{\Delta P}{\Delta I} \bigg/ \frac{hc}{q\lambda} \ , \tag{3}$$

$$\frac{\Delta P}{\Delta I} \approx \frac{1mW}{I_{1mW} - I_{th}} \ , \tag{4}$$

where c is the speed of light and h is the Planck's constant.

Starting from this part, as there are no experimental $J_{th}$ data to compare to, $I_{th}$ is used as the optimization criteria. $\eta_d$ can be calculated from $I_{th}$ and $I_{1mW}$ using Eqs. (3) and (4). The $I_{th}$ behavior can be expressed with the following equations [42]:

$$I_{th} = \frac{\frac{n}{c}\frac{1}{\Gamma \tau_p G} + n_{tr}}{\tau_s \eta_i} qWLd = J_{th}WL \tag{5}$$

$$\frac{1}{\tau_p} \cdot \frac{n}{c} = \alpha_i + \alpha_m = \alpha_i + \frac{1}{2L}\ln(\frac{1}{R_1 R_2}), \tag{6}$$

where n, G, $\tau_p$, $\tau_s$, $\eta_i$, $n_{tr}$, W, L and *d* are the group index of refraction, the proportionality factor between the gain and the carrier density, the photon life-time, the electron-hole recombination lifetime, the internal quantum efficiency, the carrier density at transparency, the cavity width, the cavity length and the cavity thickness.

Fig. 4 shows $J_{th}$, $I_{th}$ and $I_{1mW}$ at room temperature with different cavity lengths. In Fig. 4(b), we can see that there is a minimum $I_{th}$ at cavity length of 300 μm. According to Eq. (5) and (6), when L is too small, more gain and carrier density are needed due to the increased mirror loss, resulting in a larger $I_{th}$., resulting in a larger $I_{th}$. When L is too long, more current is needed to bring a large volume to transparency, which results in a monotonically increase in $I_{th}$, even in the region longer than 1 mm. The cavity length optimization slightly reduces the threshold current slightly from 790 mA to 780 mA.

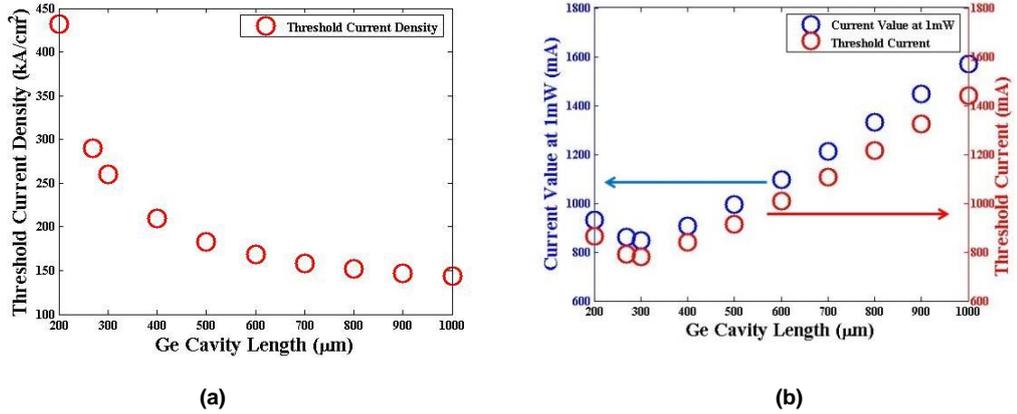

Fig.4. (a) $J_{th}$ dependence of the Ge cavity length L; (b) $I_{th}$ and $I_{1mW}$ dependence of cavity length L (Ge thickness: 200 nm, Ge width: 1μm, poly-Si thickness: 180 nm).

### 3.4 Ge cavity thickness dependence

In this section we investigated the Ge cavity thickness impact on the laser performance with the optimized cavity length at 300 μm. $J_{th}$ can be expressed as [43]

$$\eta_i \frac{J_{th}}{qd} = (An_{th} + Bn_{th}^2 + Cn_{th}^3) \tag{7}$$

$$n_{th} = n_{tr} + \frac{\alpha_i + \alpha_m}{\Gamma(d,W)G} \tag{8}$$

where *d* is the Ge cavity thickness, $\Gamma$ is the optical confinement factor, $n_{th}$ is the carrier density at threshold condition, A and C describe non-radiative recombination due to traps and surface and Auger process, respectively, B is radiative recombination coefficient, G′ is G/$\Gamma$, which is a material parameter. In Eq. (8), we can see that $n_{th}$ has thickness dependence because of $\Gamma$. In this section, because the cross-section area is constant, $J_{th}$ and $I_{th}$ share the same trend.

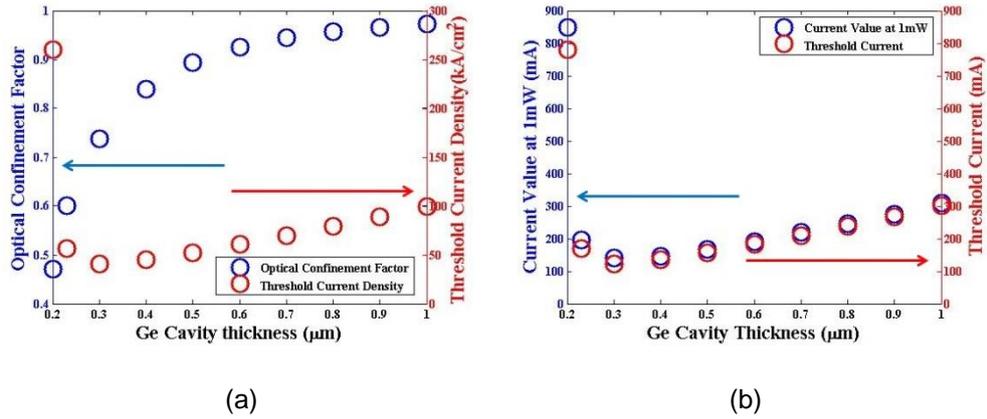

(a)                          (b)

Fig. 5. (a) $\Gamma$ and $J_{th}$ dependence of the Ge cavity thickness d in the range of 0.2 to 1 μm; (b) the dependence of $I_{th}$ and $I_{1mW}$ on the Ge cavity thickness d (Ge length: 300 μm, Ge width: 1μm, poly-Si thickness: 180 nm).

Fig. 5 (b) shows the calculated $I_{th}$ as a function of Ge cavity thickness d from 0.2 to 1 μm. We can see that the lowest $I_{th}$ is obtained at $d = 0.3$ μm. When $d$ is too small, an increasing fraction of the optical field is outside the active region due to the smaller optical confinement factor $\Gamma$. In Fig. 5(a), we can see that in thinner thickness, the $J_{th}$ and $I_{th}$ are significantly affected by $\Gamma$. On the thicker end when $\Gamma$ is approaching 1, $I_{th}$ is linear with $d$. The laser will also be more efficient with a thicker Ge cavity. The optimum $I_{th}$ is around $d = 0.3$ μm, where the emission wavelength will also red shift with the increasing thickness to $\lambda=1700$ nm [21]. The cavity thickness optimization reduces the threshold current from 780 mA to 123 mA.

### 3.5 Ge cavity width dependence

With the optimized Ge cavity length of 300 μm and thickness of 0.3 μm, next, we investigated the cavity width's influence on the Ge laser. Fig. 6 (a) shows $J_{th}$ vs. different cavity width W with the optimized Ge length, thickness and the fixed poly-Si thickness of 180 nm, respectively. According to Eq. (8), with the decreasing cavity width, $J_{th}$ became higher due to the decrease of $\Gamma$, which can be also seen in Fig. 6(a).

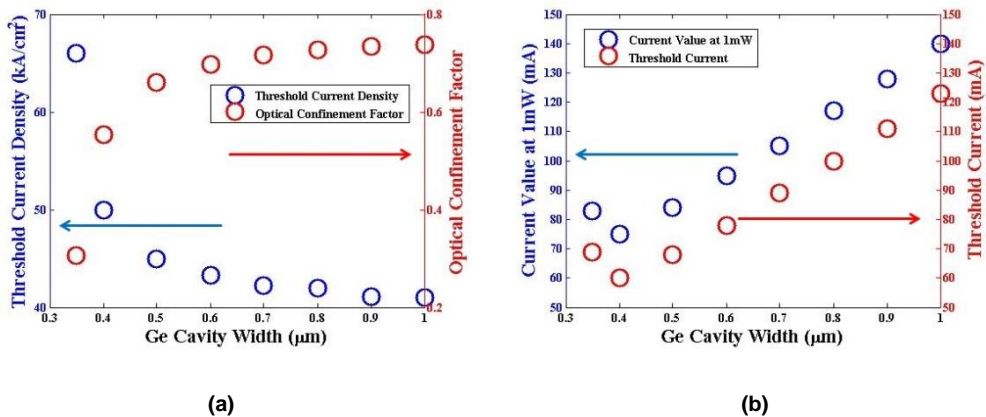

(a)                          (b)

Fig.6. (a) Optical confinement factor and Threshold current density dependence of the Ge cavity width in the range of 0.3~1 μm; (b) Threshold current and Current value at output light power of 1mW between the Ge cavity width varied from 0.3~1μm (Ge length: 300 μm, Ge thickness: 0.3 μm, poly-Si thickness: 180 nm).

When considering the $I_{th}$ ($I_{th}=J_{th}*W*L$), it has a different W dependence. A wider laser requires more injection current to reach the onset of the lasing action. We can see that in Fig. 6(b) that between 0.3 μm to 1 μm, the lowest threshold current value is at Ge cavity width of 0.4 μm. In Fig. 6(b), $I_{1mW}$ is also shown. The similar trend between $I_{th}$ and $I_{1mW}$ for W > 0.4 μm is expected as $\Gamma$ and thus $\eta_d$ is not sensitive to Ge width in this range, due to the large difference between the refractive index in Ge and $SiO_2$. The cavity width optimization reduces the threshold current from 123mA to 60 mA.

### 3.6 Poly-Si cladding thickness dependence

Now that we have determined the Ge cavity size (thickness, width and length are 0.3 μm, 0.4 μm and 300 μm, respectively), we can optimize the thickness of the poly-Si cladding layer. The high $J_{th}$ 280 kA/cm$^2$ in the experimental data is partially due to the lossy metal contact, where the poly-Si thickness was 180 nm. If the metal contact is moved further away from the Ge cavity, the losses would decrease monotonically. The simulation results are shown in Fig. 7. We chose the optimal thickness of the poly-Si layer to be 0.6 μm, above which the extra decrease in $I_{th}$ is negligible. It can also be seen that the difference in $I_{th}$ and $I_{1mW}$ shrinks in each optimization step, showing an increase in $\eta_d$, which is 22.5% after the structure optimizations. The poly-Si cladding thickness optimization reduces the threshold current from 60 mA to 36 mA.

After all the structure optimization steps, compared to the original $I_{th}$ =790 mA and $\eta_d$ = 2.0% , there is a 22-fold reduction in $I_{th}$ and a 11-fold increase in $\eta_d$. The improvements are mainly from the better optical confinement factor and less optical loss due to the metal contacts.

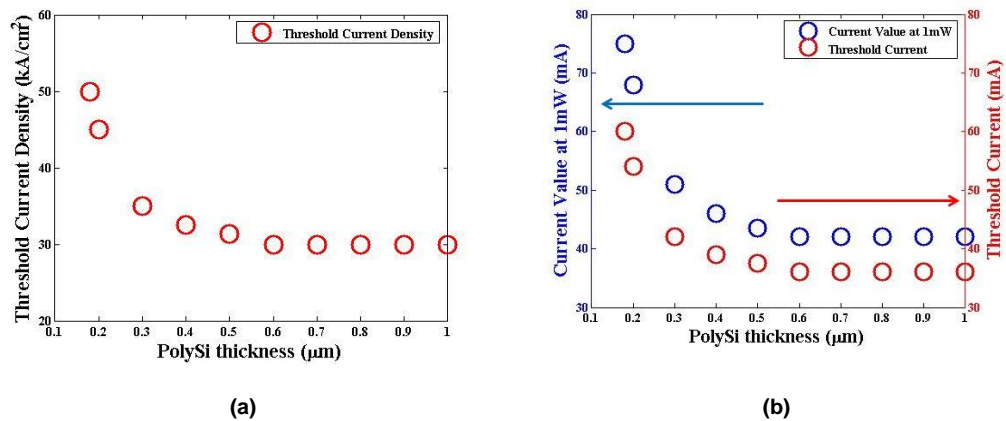

Fig.7. (a) $J_{th}$ dependence of the poly-Si thickness in the range of 0.18 to 1 μm; (b) $I_{th}$ and $I_{1mW}$ between the poly-Si thickness in the range of 0.18 to 1 μm (Ge length: 300 μm, Ge thickness: 0.3 μm, Ge width: 0.4 μm).

### 3.7 Defect limited minority carrier lifetime dependence

With the optimized Ge cavity size (thickness 0.3 μm, width 0.4 μm, length 300 μm) and the poly-Si cladding layer thickness (0.6 μm), our models predict that $J_{th}$ of 30 kA/cm$^2$, $I_{th}$ of 36 mA, $I_{1mW}$ of 42 mA and $V_{1mW}$ of 0.84V. The wall-plug efficiency $\eta_{wp}$ of 2.8% can be achieved with 1 ns defect limited carrier lifetime. As previously discussed in 3.2, 1 ns lifetime is a conservative estimation. For lasers in Ref. [21], carriers were concentrated at the p-n junction, which was designed to be away from the highly defected Ge/Si interfaces. Technically, it is feasible to obtain Ge layers with better quality and longer carrier lifetimes by approaches of Ge growth on a GOI substrate [44] or direct wafer bonding and chemical mechanical polishing (CMP) [45]. Minority carrier lifetimes of 5.3 and 3.12 ns have been achieved respectively by the above approaches [44, 45]. To investigate the impact of the carrier lifetime, we performed simulations with various carrier lifetimes up to 100 ns with all other conditions unchanged. It can be seen that $I_{th}$ and $\eta_{wp}$ improve with the carrier lifetime very rapidly showing the essential role of minority carrier lifetime on the threshold current reduction and the efficiency enhancement, which is consistent with the recent work by D. S. Sukhdeo et al [46]. With a 10 ns lifetime, $I_{1mW}$ of 6.3 mA and $\eta_{wp}$ of 10.9% can be achieved (Fig. 8). With a 100 ns lifetime, the laser could achieve a $J_{th}$ of 2.8 kA/cm$^2$, $I_{th}$ of 3.3 mA, $I_{1mW}$ of 9 mA, $V_{1mW}$ of 0.76V, $\eta_d$ of 23.6%  and $\eta_{wp}$ of 14.6% (Fig. 8). This makes the Ge laser performance much closer to that of common III-V semiconductor lasers. The L-I and contact voltage vs. contact current (V-I) characteristics with 1 and 100 ns carrier lifetimes are shown in Fig. 9.

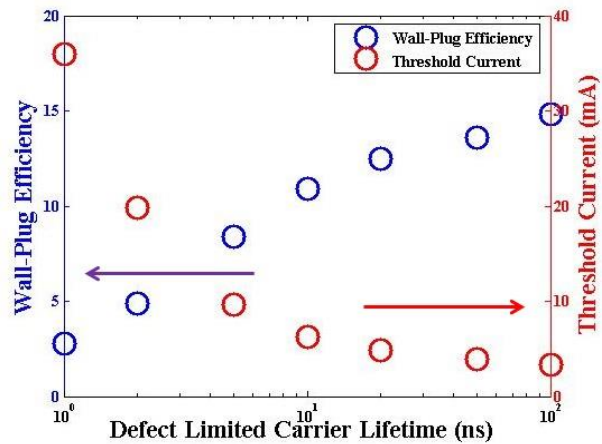

Fig.8. Wall-plug efficiency and threshold current of the Ge laser with various defect limited carrier lifetimes.

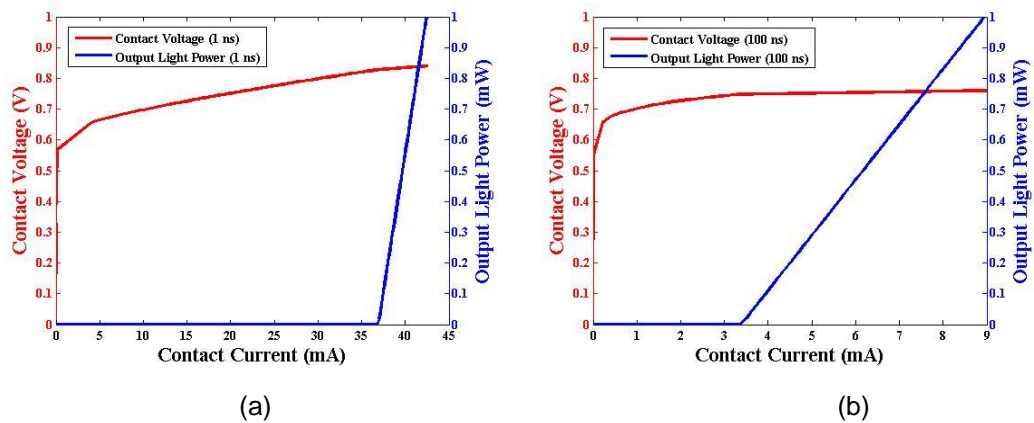

(a)                                                                                  (b)

Fig.9. Optimized L-I and V-I characteristics with different defect limited carrier lifetime 1 ns (a) and 100 ns (b) (Ge: thickness 0.3 μm, width 0.4μm, length 300μm, poly-Si: cladding thickness 0.6 μm)

## 4. Conclusions

Physical models of Ge energy band structure and material loss were implemented in LASTIP$^{TM}$, a 2D simulation tool for edge emitting laser diodes. This model is able to match experimental data available. Important design parameters of a Fabry-Perot Ge laser, such as the cavity length, thickness, width, polycrystalline Si cladding layer thickness were studied and optimized. The laser structure optimizations alone were shown to reduce $I_{th}$ by 22-fold and increase $\eta_d$ by 11 times. The simulations also showed that improving the defect limited carrier lifetime is critical for achieving an efficient and low-threshold Ge laser. With the optimized structure design (300 μm for the cavity length, 0.4 μm for the cavity width, 0.3 μm for the cavity thickness, and 0.6 μm for the polycrystalline Si cladding layer thickness) and a defect limited carrier lifetime of 100 ns, a wall-plug efficiency of 14.6% at 1mW output is predicted, where $J_{th}$ of 2.8 kA/cm$^2$, $I_{th}$ of 3.3 mA, $I_{1mW}$ of 9 mA, and $\eta_d$ of 23.6% can be achieved. These are tremendous improvements from the available experimental values at 280 kA/cm$^2$, 756 mA, 837 mA and 1.9%, respectively.

Considering that common commercial compound semiconductor lasers have wall-plug efficiencies in the range of 20-30%, that of the optimized Ge laser is about half of that. Strain and doping optimization, quantum-well structures and material quality improvements, although outside the scope of this work, are all very important to improve the Ge laser performance. With future efforts in all other design and processing aspects, we are optimistic to expect Ge lasers with much better performance realized in the near future.

**Acknowledgments**

This work was supported by the Oversea Study Program of Guangzhou Elite Project and Crosslight Software Inc.. The authors would like to thank Mr. Michel Lestrade from Crosslight Software for his assistance in LASTIP$^{TM}$ usage, Mr. Yiheng Lin from the Department of Materials Engineering at the University of British Columbia and Prof. Jifeng Liu from Thayer School of Engineering at Dartmouth College for helpful discussions.